\documentclass[pra,aps,showpacs,floatfix,nofootinbib,twocolumn,letterpaper,reprint]{revtex4-2}
\usepackage{epsfig}
\usepackage{amssymb}
\usepackage{amsmath}
\usepackage{amsfonts}
\usepackage{color,graphicx}
\usepackage{float}
\usepackage{comment}
\usepackage{bm}
\usepackage[pdftex, pdfborder= 0 0 0, citecolor=blue, urlcolor=blue, linkcolor=blue, colorlinks=true, bookmarksopen=true]{hyperref}

\newcommand{\ssec}[1]{\emph{#1}.---}
\DeclareMathOperator{\Tr}{Tr}

\definecolor{violet}{RGB}{100,0,200}

\begin{document}
\title{Precision Thermodynamics of the Fermi polaron at strong coupling}
\author{S. Ramachandran$^1$}
\author{S. Jensen$^2$}
\author{Y.  Alhassid$^1$}
\affiliation{$^1$Center for Theoretical Physics, Sloane Physics Laboratory, Yale University, New Haven, Connecticut 06520, USA
\\
$^2$Department of Physics, University of Illinois at Urbana-Champaign, Urbana, Illinois 61801, USA}

\begin{abstract}
The Fermi polaron problem, which describes a mobile impurity that interacts with a spin-polarized Fermi sea, is a paradigmatic system in quantum many-body physics and has been challenging to address quantitatively in its strong coupling regime. We present the first controlled thermodynamic calculations for the Fermi polaron at strong coupling using finite-temperature auxiliary-field quantum Monte Carlo (AFMC) methods in the framework of the canonical ensemble.  Modeled as a spin-imbalanced system, the Fermi polaron  has a Monte Carlo sign problem, but we show that it is moderate over a wide range of temperatures and coupling strengths beyond the unitary limit of the BCS-BEC crossover. We calculate the contact, a quantity which measures the strength of the short-range correlations, as a function of temperature at unitarity and as a function of the coupling strength at fixed temperature and find good agreement with a variational approach based on one particle-hole excitation of the Fermi sea. We compare our results for the contact with recent experiments and find good agreement at unitarity (within error bars) but  discrepancies away from unitarity on the BEC side of the crossover.  We also calculate the thermal energy gap at unitarity as a function of temperature.  
\end{abstract}

\maketitle{}

\section{Introduction} 

The Fermi polaron describes a mobile impurity interacting with a spin-polarized medium.  Initially investigated by Landau and Pekar as a simplified model for electrons on the lattice~\cite{Landau1948}, the polaron has since been studied across energy scales, including in solid state materials~\cite{Sidler2017,Tan2020,Lindemann1983}, superfluid Helium~\cite{Kondo1983}, and nucleon systems~\cite{Kutschera1993,Nakano2020,Vidana2021}. The polaron straddles two key problems in many-body physics: the impurity physics of minority particles in a medium and the quasiparticle physics of bare particles dressed by interactions. The Fermi polaron describes the interaction of an impurity with a fully polarized Fermi sea; the analogous problem for bosons has been the subject of much interest~\cite{Catani2012,Hu2016,Jorgensen2016,Camargo2018,Yan2020}. Ultracold Fermi gases have recently been used to realize a controllable Fermi polaron in the context of strongly correlated spin-imbalanced Fermi liquids~\cite{Yan2019,Ness2020,Nascimbene2009,Cetina2016}. These ultracold gases provide a potentially ideal environment to study the competing orders of superfluidity and magnetism~\cite{Navon2010,Lobo2006}. Understanding the properties of the polaron's excitations is crucial for understanding the superfluidity as well as for identifying the time scales of magnetic stability in these systems~\cite{Cui2010, Scazza2017, Darkwah2019, Kohstall2012}. 

In the BCS-BEC crossover, the polaron is studied as a function of $(k_F a)^{-1}$, where $k_F$ is the Fermi momentum of the medium and $a$ is the scattering length characterizing the short-range $s$-wave interaction between the impurity and the medium~\cite{Schirotzek2009,Cetina2015}. Quasiparticle excitations include the attractive polaron, a dressed quasiparticle; the repulsive polaron, a semistable quasiparticle~\cite{Massignan2011,Schmidt2011,Kohstall2012,Scazza2017}; and the dimer molecule. In the unitary limit of $(k_F a)^{-1}=0$ (i.e., infinite scattering length), there is a crossover from a classical gas at high temperatures to the Fermi polaron at low temperatures~\cite{Yan2019}. Additionally, at low temperatures, a first-order phase transition is predicted with increasing $(k_F a)^{-1}$  from the an attractive polaron to a dimer molecule~\cite{Parish2021}. Estimates for the critical coupling strength range from $(k_F a) ^{-1} \approx 0.9$~\cite{Schmidt2011} to $(k_F a) ^{-1} \approx 1.3$~\cite{Ness2020}. Both the polaron-molecule transition~\cite{Ness2020} and the thermodynamics of the Fermi polaron at unitarity~\cite{Yan2019} were investigated through ultracold atoms experiments.

Theoretical studies have explored the ground-state properties of the system using diffusion Monte Carlo~\cite{Pessoa2021, Lobo2006}, diagrammatic Monte Carlo~\cite{Prokof'ev2008, Goulko2016}, functional renormalization group methods~\cite{Schmidt2011}, and the one particle-hole approximation~\cite{Chevy2006, Punk2009}. At finite temperature, theoretical calculations have been performed using the one-particle-hole approximation~\cite{Massignan2008,Tajima2019,Liu2019,Liu2020B,Liu2020A,Parish2021,Mulkerin2019,Hu2018,Hu2022}.  However, there have been no controlled calculations at finite temperature, probably due to a Monte Carlo sign problem for spin-imbalanced systems.

We present the first controlled finite-temperature calculations of the Fermi polaron thermodynamics at strong coupling using the auxiliary-field quantum Monte Carlo (AFMC) method in the framework of the canonical ensemble, for which the numbers of both the majority and minority particles are fixed. In particular, the canonical-ensemble AFMC allows for an accurate and precise investigation of a single impurity; in the grand-canonical ensemble, although large spin imbalances are accessible, it is difficult to isolate the case of a single impurity. 

Here, we first explore the behavior of the Monte Carlo sign in the strongly interacting regime and demonstrate that the effects of the sign problem are minimal over a wide range of temperatures and coupling strengths. We then calculate Tan's contact, a quantity which measures the strength of the short-range correlations, over this region and compare the results with recent experiments~\cite{Yan2019, Ness2020} and theory~\cite{Parish2021, Liu2020A, Pessoa2021}.

We further calculate the thermal energy gap, defined as the difference between the thermal energies of the interacting and non-interacting systems. Our low-temperature results are consistent with the ground-state energy gap $\Delta E = -0.61 E_F$ provided by Chevy's ansatz~\cite{Chevy2006}. 

\section{Finite-temperature canonical ensemble AFMC} 

To study the Fermi polaron, we consider a system of spin-$1/2$ fermions at extreme spin polarization, with $N_\uparrow$ spin-up particles interacting with a single spin-down impurity (i.e., $N_\downarrow=1$).  We model the short-range interaction between the impurity and the polarized Fermi sea by a contact interaction tuned by the two-particle scattering length $a$. The continuum Hamiltonian of this system is 
\begin{equation}\label{eq:continuumhamiltonian}
\begin{split}
\hat{H} = \sum_{s_z} \int d^3 \mathbf{r} \ \hat{\psi}_{s_z} ^\dagger (\mathbf{r}) \left ( -\frac{\hbar^2 \nabla^2 _{\mathbf{r}}}{2m} \right ) \hat{\psi}_{s_z} (\mathbf{r})
\\
+ V_0 \int d^3 \mathbf{r} \ \hat{\psi}_{\uparrow} ^\dagger (\mathbf{r}) \hat{\psi}_{\downarrow} ^\dagger (\mathbf{r}) \hat{\psi}_{\downarrow} (\mathbf{r}) \hat{\psi}_{\uparrow} (\mathbf{r}) \;,
\end{split}
\end{equation}
where $\hat{\psi}_{s_z}^\dagger (\mathbf{r}) $ is the creation operator for a particle with spin $s_z$ at position $\mathbf{r}$, $V_0<0$ is the bare interaction strength, and $-\hbar^2 \nabla^2 _{\mathbf{r}}/2m$ is the single-particle kinetic energy operator. We use  the canonical ensemble, in which both  $N_\uparrow$ and $N_\downarrow$ are fixed and do not fluctuate. This allows us to consider the case of a single impurity $N_\downarrow=1$. In contrast, in the grand-canonical ensemble, while large values of spin imbalance $N_\uparrow / N_\downarrow >> 1$ are accessible, it is difficult to isolate the case of a single impurity. To describe the thermodynamics of this continuum system, we consider a discrete lattice  of $N_L ^3$ points with lattice spacing of $\delta x$ and periodic boundary conditions. The lattice Hamiltonian is given by
\begin{equation}\label{eq:latticehamiltonian}
\hat{H} = \sum_{\mathbf{k}, s_z} \epsilon_{\mathbf{k}} \hat{a}^\dagger _{\mathbf{k}, s_z} \hat{a} _{\mathbf{k}, s_z} + g \sum_{\mathbf{x}_i} \hat{n}_{\mathbf{x}_i , \uparrow} \hat{n}_{\mathbf{x}_i , \downarrow} \;,
\end{equation}
where $\epsilon_{\mathbf{k}} =\hbar^2 k^2/2m$.  The interaction strength $g = \frac{V_0}{(\delta x )^3}$ is determined so as to reproduce the given scattering length $a$ on the lattice. This leads to
\begin{equation}\label{eq:renormalizedInteraction}
\frac{1}{V_0} = \frac{m}{4\pi\hbar^2 a} - \int_{B} \frac{d^3\mathbf{k}}{(2\pi)^3 2\epsilon_{k}} \;,
\end{equation}
where the integral is taken over the entire first Brillouin zone $B$. 

In the AFMC approach, we divide the imaginary-time interval $\beta$ into $N_\tau$ discrete time slices of width $\Delta\beta = \beta/N_\tau$ and apply a symmetric Trotter-Suzuki decomposition to the thermal propagator
\begin{equation}\label{eq:trotter}
\begin{split}
e^{-\beta\hat{H}} & = (e^{-\Delta\beta\hat{H}})^{N_{\tau}} \\
 & = \prod_{i=1}^{N_{\tau}}e^{-\frac{\Delta\beta}{2}\hat{K}}e^{-\Delta\beta\hat{V}}e^{-\frac{\Delta\beta}{2}\hat{K}} + O(\Delta\beta^2)\;.
\end{split}
\end{equation}
Here $\hat{K} = \sum_{\mathbf{k},s_z} \epsilon_{\mathbf{k}} \hat{a}^\dagger _{\mathbf{k},s_z} \hat{a}_{\mathbf{k},s_z} - g (\hat{N}_\uparrow + \hat{N}_\downarrow)/2$ is a one-body operator and $\hat{V} = g/2 \sum_{\mathbf{x}_i} \hat{n}_{\mathbf{x}_i} ^2$ with $\hat{n}_{\mathbf{x}_i} = \hat{n}_{\mathbf{x}_i , \uparrow} + \hat{n}_{\mathbf{x}_i , \downarrow}$ is a two-body interaction. A Hubbard-Stratonovich transformation is applied to linearize $\hat V$, introducing  auxiliary fields for each lattice site $\mathbf{x}_i$ and time slice $\tau_n$. For a single lattice site and time slice this is described by the HS transformation
\begin{equation}\label{eq:HSlocal}
e^{-\Delta\beta g \hat{n}_{\mathbf{x}_i} ^2 /2} = \sqrt{\frac{\Delta\beta |g|}{2\pi}} \int_{-\infty} ^\infty d\sigma_{\mathbf{x}_i} e^{-\Delta\beta |g| \sigma_{\mathbf{x}_i} ^2 / 2} e^{-\Delta\beta g \sigma_{\mathbf{x}_i} \hat{n}_{\mathbf{x_i}}}\;.
\end{equation}

\begin{figure*}[bth]
 \includegraphics[scale=0.5]{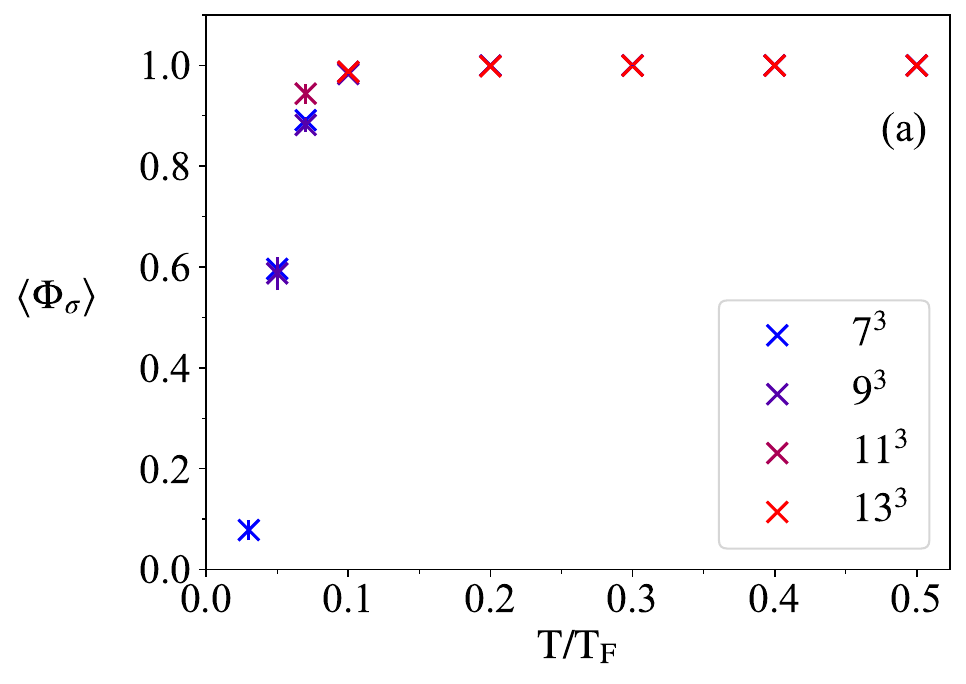}
 \includegraphics[scale=0.5]{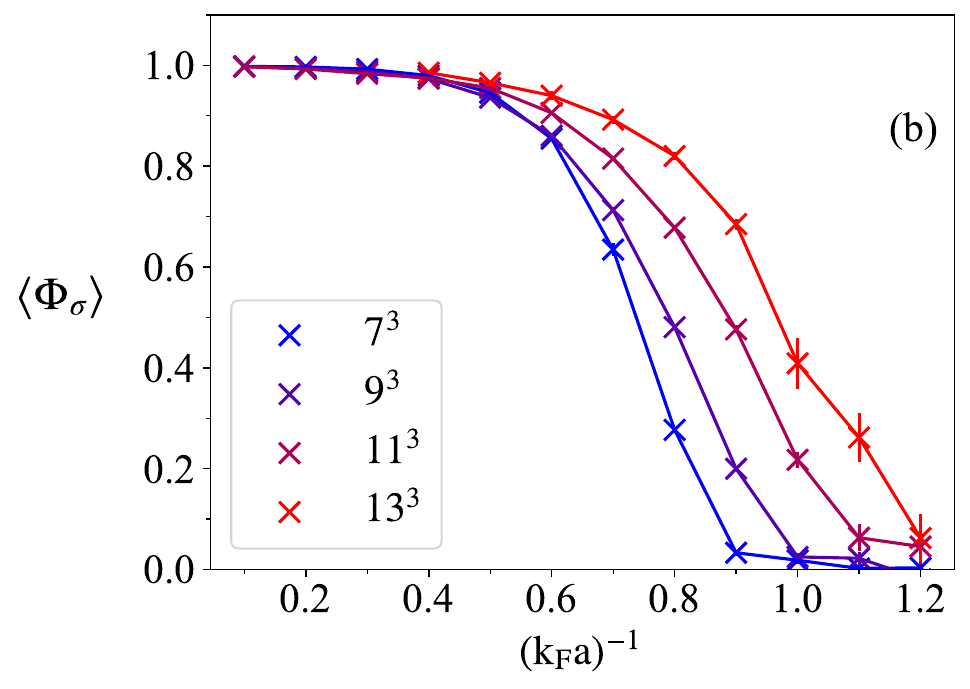}
 \caption{ (a)  Average Monte Carlo sign as a function of temperature at unitarity for 20+1 particles on  lattices of size $7^3$, $9^3$, $11^3$, and $13^3$. 
  (b)  As in (a) but for the average Monte Carlo sign as a function of $(k_F a)^{-1}$ at $T=0.2\ T_F$.}
\label{fig:Sign}
\end{figure*}

We discretize the path integral over continuous fields using a 3-point Gaussian quadrature~\cite{Koonin1997} which introduces an overall systematic error of $\mathcal{O} ( (\Delta\beta)^2 )$, of the same order as the error introduced by the symmetric Trotter-Suzuki decomposition. The thermal expectation of an observable $\hat O$ is given by
\begin{equation}\label{eq:path_integral}
\langle \hat{O} \rangle=\frac{\mathrm{Tr}(\hat{O}e^{-\beta \hat{H}})}{\mathrm{Tr}(e^{-\beta \hat{H}})}=\frac{\int D[\sigma]G_{\sigma} \langle \hat{O} \rangle _{\sigma}\mathrm{Tr} \hat{U}_{\sigma} }{\int D[\sigma]G_{\sigma}\mathrm{Tr} \hat{U}_{\sigma}} \;,
\end{equation}
where $\langle \hat{O} \rangle _{\sigma}$ is the expectation value of the observable and $\mathrm{Tr}(\hat{U}_{\sigma})$ is the partition function for a given field configuration $\sigma$. Here $D[\sigma]$ is an integration measure and $G_\sigma$ is a Gaussian weight given by
\begin{equation}\label{eq:HSparts}
\begin{split}
D[\mathbf{\sigma}] = \prod_{n=1} ^{N_\tau}  \prod_{\mathbf{x}_i} \left ( \sqrt{\frac{\Delta\beta |g|}{2\pi}} \right ) \textup{d}\sigma_{\mathbf{x}_i} (\tau_n)
\\
G_\mathbf{\sigma} = e^{-\Delta\beta |g| \sum_{i, n} \sigma^2 _{\mathbf{x}_i} (\tau_n) / 2} \;.
\end{split}
\end{equation}
The expectation value in Eq.~(\ref{eq:path_integral}) is evaluated using Monte Carlo methods by averaging over $\sigma$-field configurations that are sampled by the Metropolis-Hastings-Rosenbluth algorithm according to the weight function $W_\sigma=G_\sigma | \Tr{\hat{U}_\sigma} |$. 
Since $U_\sigma$ is a one-body propagator, the integrand in Eq.(\ref{eq:path_integral}) can be calculated by linear algebra in the single-particle space. For example
\begin{equation}\label{eq:tracedeterminant}
\Tr  \hat{U}_\sigma  = \det (1 + \mathbf{U}_\sigma) \;,
\end{equation}
where $\mathbf{U}_\sigma$ is the matrix representing  the thermal propagator $U_\sigma$ in the single-particle space. We use a canonical ensemble formulation of AFMC in which the traces are taken over fixed particle number. This is achieved through an exact particle-number projection described by a finite Fourier sum
\begin{equation}\label{eq:particlenumberprojection}
\hat{P}_{N_{s_z}} = \frac{e^{-\beta \mu N_{s_z}}}{N_s} \sum_{m = 1} ^{N_s} e^{-i \varphi_m N_{s_z}} e^{\left ( \beta \mu + i \varphi_m \right ) \hat{N}_{s_z}} \;.
\end{equation}
where $N_s=N_L^3$ is the number of single-particle states for a given spin value $s_z$ and $\varphi_m= 2\pi m/N_s$ ($m=1,\ldots, N_s$) are discrete quadrature points. The real chemical potential $\mu$ is chosen to give the correct average number of particles to guarantee the numerical stability of the Fourier sum.   In the canonical formulation, we replace the grand-canonical traces $\Tr \hat X$ with canonical traces $\Tr_{N_\uparrow, N_\downarrow} \hat{X} = \Tr (\hat{P}_{N_{\uparrow}} \hat{P}_{N_{\downarrow}} \hat{X})$. 

\section{Monte Carlo sign function}

The one-body propagator factorizes $\hat U_\sigma = \hat U^\uparrow_\sigma  \hat U^\downarrow_\sigma$ for each auxiliary-field configuration $\sigma$ and so does the projected 
partition function
\begin{equation}
\Tr_{N_\uparrow, N_\downarrow} \hat  U_\sigma = (\Tr_{N_\uparrow} \hat  U^\uparrow_\sigma) ( \Tr_{N_\downarrow} \hat  U^\downarrow_\sigma) \;.
\end{equation}
  
 For an attractive contact interaction (i.e., $g <0$), the propagator $\hat U_\sigma$ is invariant under time reversal and therefore the spin-up and spin-down projected partition functions, $\Tr_{N_\uparrow} \hat  U_\sigma$ and $\Tr_{N_\downarrow} \hat  U_\sigma$, are both real numbers.  The Monte Carlo sign function $\Phi_\sigma$ is then the product of the signs of the projected spin-up and spin-down partition functions. For the spin-balanced case $N_\uparrow=N_\downarrow$, the two projected partitions are equal and $\Tr_{N_\uparrow, N_\downarrow} \hat  U_\sigma = (\Tr_{N_\uparrow}\hat  U^\uparrow_\sigma)^2 > 0$. In that case the Monte Carlo sign function is 1 for each sample $\sigma$ and there is no sign problem. However, for the spin-imbalanced case $N_\uparrow \neq N_\downarrow$, the sign can be negative for some of the samples and in general one expects to have a sign problem.

The Monte Carlo sign problem is usually considered a major obstacle in performing precision calculations in spin-imbalanced systems. The polaron problem describes an extreme spin imbalance case with $N_\downarrow=1$. We investigated the average Monte Carlo sign $\langle \Phi_\sigma \rangle$ in the parameter space of the polaron, i.e., as a function of temperature and coupling strength $(k_F a)^{-1}$.

In Fig.~\ref{fig:Sign}(a) we show the average Monte Carlo sign as a function of temperature at unitarity $(k_F a)^{-1}=0$ for a polaron problem with $20$ spin-up  particles and $1$ spin-down particle (the impurity), which we refer to as the $20+1$ system.  Results are shown for lattice sizes of $N_L^3 =7^3, 9^3, 11^3$ and $13^3$.  The average sign is independent of the lattice size and  is close to 1 for $T>0.1\ T_F$. It declines rapidly below $T ~\sim 0.1\,T_F$, and at $T \sim 0.05 T_F$ the average sign is $\sim 0.6$.  Most experiments access the regime $T>0.1\ T_F$, where the Monte Carlo sign is rather good.

In Fig.~\ref{fig:Sign}(b), we show the average Monte Carlo sign as a function of the coupling $(k_F a)^{-1}$ for the $20+1$ system at a constant temperature of $T=0.2\ T_F$, a low but experimentally accessible~\cite{Yan2019,Ness2020} temperature. We find that the Monte Carlo sign is still good as we enter the BEC regime but it starts to decline rapidly above $(k_F a)^{-1} \approx 0.7$ for the small lattice sizes.  At $(k_F a)^{-1} \approx 0.9$ the sign is $\sim 0.2$ for the $9^3$ lattice, for which meaningful  Monte Carlo calculations are still feasible. This allows us to perform calculations up to the polaron-molecule transition point, but does not allow us to probe the transition itself.  We also observe that the sign improves with increasing lattice size.\\

\section{Results} 

In this section we present finite-temperature AFMC results for the Fermi polaron problem.  We carried out calculations for $N_{\uparrow}=7$ particles with a single impurity  (7+1 system) and for $N_{\uparrow}=20$ particles with a single impurity (20+1 system). We find convergence in $N_{\uparrow}$, implying that we have reached the thermodynamic limit. The convergence to the thermodynamic limit at relatively small $N_{\uparrow}$, reflects the rapid convergence of the spin-polarized majority to a Fermi sea; convergence to a Fermi sea has been experimentally demonstrated using as few as 5 particles~\cite{Wenz2013}.

\subsection{Contact}

Tan's contact $C$ is a fundamental thermodynamic property of quantum many-body systems with short-range interactions, which describes the short-range correlations between particles of opposite spin.  It is defined by
\begin{equation}\label{eq:contactdefinition}
\int d^3 \mathbf{R} \ g^{(2)} _{\uparrow\downarrow} (\mathbf{R} + \mathbf{r} / 2, \mathbf{R} - \mathbf{r} / 2) \underset{r\rightarrow 0}{\sim}\frac{C}{4\pi r^2} \;,
\end{equation}
where $g^{(2)} _{\uparrow\downarrow} (\mathbf{r}_\uparrow, \mathbf{r}_\downarrow) = \langle \hat{n}_\uparrow (\mathbf{r}_\uparrow) \hat{n}_\downarrow (\mathbf{r}_\downarrow) \rangle$ is the two-particle correlation function, and $\hat n_{s_z}(\mathbf{r})$ is the density of particles with spin $s_z=\uparrow,\downarrow$ at position $\mathbf r$. The contact appears in several relations known as Tan's relations~\cite{Tan2008a,Tan2008b,Tan2008c}, an example being the tail of the momentum distribution, $n_{s_z} (\mathbf{k}) \underset{k\rightarrow 0}{\sim}C/k^4$. The contact is also related to partial derivatives of the energy and free energy with respect to the inverse scattering length at constant entropy $S$  and  constant temperature, respectively
\begin{equation}\label{eq:contactderivatives}
\begin{split}
C = \frac{4\pi m}{\hbar^2} \left. \frac{\partial E}{\partial (-1/a)} \right | _S \;,
\\
C = \frac{4\pi m}{\hbar^2} \left. \frac{\partial F}{\partial (-1/a)} \right | _T \;.
\end{split}
\end{equation}
 In the lattice formulation, the second relation leads to  
\begin{equation}\label{eq:contactV}
C = \frac{m^2 V_0 \langle \hat{V} \rangle }{\hbar^4} \;,
\end{equation}
where $\langle \hat{V} \rangle$ is the thermal expectation value of the potential energy operator $\hat V = g \sum_{\mathbf{x}_i} \hat{n}_{\mathbf{x}_i , \uparrow} \hat{n}_{\mathbf{x}_i , \downarrow}$ in the lattice model. 

\subsubsection{Contact at unitarity versus temperature}

\begin{figure}
    \centering
    \hspace*{-0.7cm}
    \includegraphics[scale=0.6]{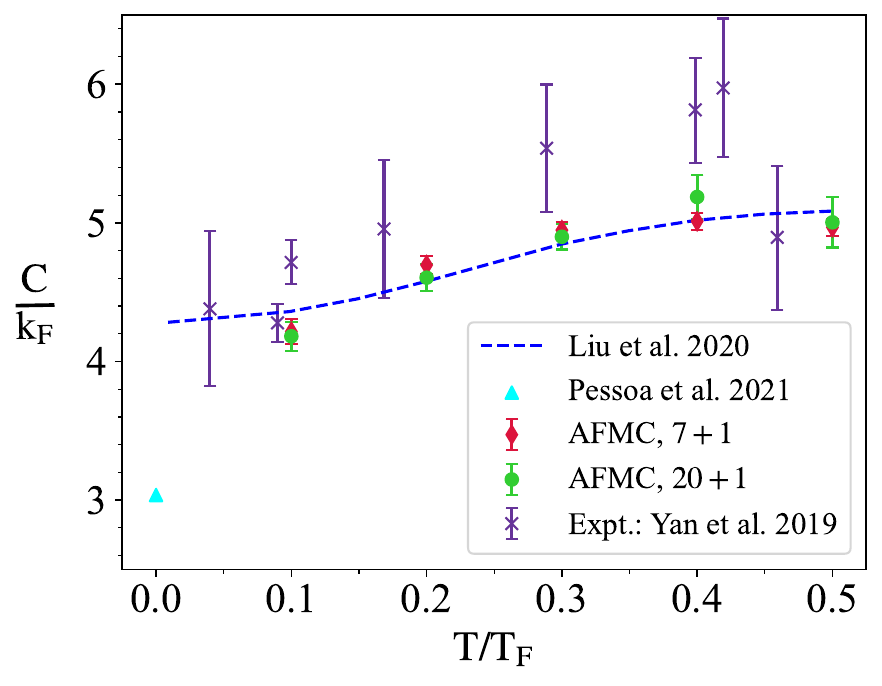}
    \caption{Contact $C$ (in units of $k_F$) vs.~temperature $T$ (in units of the Fermi temperature $T_F$) at unitarity.  The AFMC results for the $7+1$ system (red diamonds)  and the $20+1$ system (green circles) are compared the variational results of Ref.~\cite{Liu2020A} (dashed blue line).  The $T=0$ diffusion Monte Carlo result of Ref.~\cite{Pessoa2021} is shown by the blue triangle. We also show the experimental results of Ref.~\cite{Yan2019} (purple x's with error bars).}
 \label{fig:CvsT}
\end{figure}

In Fig.~\ref{fig:CvsT} we show the contact at unitarity ($(k_F a)^{-1} = 0$) as a function of temperature for the 7+1 (red diamonds) and 20+1 (green circles) systems.  We eliminate systematic errors by extrapolating to continuous time $\Delta \beta \rightarrow 0$ and taking the continuum limit $\nu \rightarrow 0$,  where $\nu=N_\uparrow/N_L^3$ is the filling factor. These extrapolations are demonstrated in Appendix A. 

Our results are in close agreement with the variational result of Ref.~\cite{Liu2020A}, which are based on one particle-hole excitation of the Fermi sea.    However, the ground-state diffusion Monte Carlo result of Ref.~\cite{Pessoa2021} is significantly smaller that the low-temperature  AFMC results.

 The monotonic increase of the contact at unitarity as a function of $T$ below $0.5\, T_F$ reflects  the increased population of the excited molecular states at higher temperatures~\cite{Liu2020A,Parish2021,Schmidt2011}.  We also find overall agreement of our results with the experimental results of Ref.~\cite{Yan2019} when considering the large experimental errors bars.

\begin{figure*}[bth]
    \includegraphics[scale=0.55]{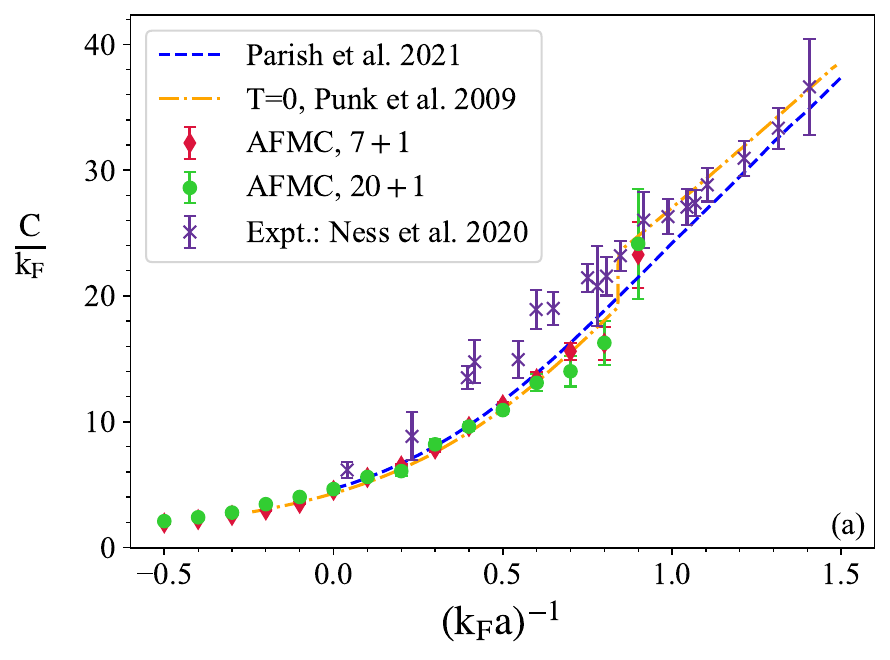}
    \includegraphics[scale=0.55]{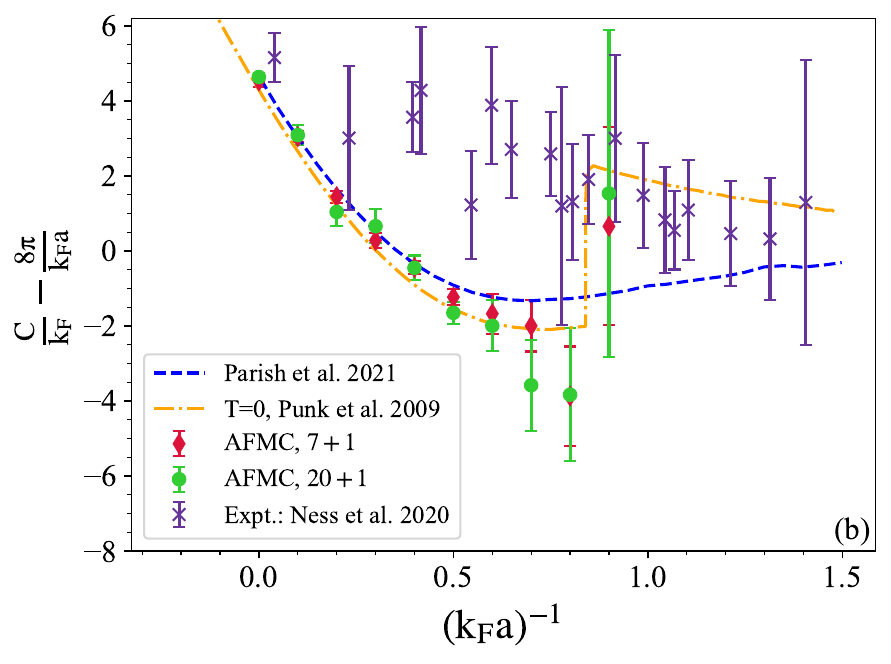}
    \caption{(a) Contact $C$ (in units of $k_F$) as a function of coupling strength $1/(k_F a)$ at $T = 0.2\,T_F$.  The AFMC results for the $7+1$ system (red diamonds)  and the $20+1$ system (green circles) are compared the variational results of Ref.~\cite{Parish2021} (dashed blue line), the $T=0$ functional renormalization group results of Ref.~\cite{Punk2009} (dash-dotted orange line), and the experimental results of Ref.~\cite{Ness2020} (purple x's with error bars).
   (b) As in panel (a) but for the contact $C/k_F$ shifted by the contribution $8\pi/ k_Fa$ of the two-particle binding energy.}
 \label{fig:Cvsa}
\end{figure*}

\subsubsection{Contact versus coupling strength}

In Fig.~\ref{fig:Cvsa}(a) we show the contact as a function of the coupling strength $(k_F a)^{-1}$ at a constant temperature of $T=0.2\,T_F$. Our canonical-ensemble AFMC results for the 7+1 (red diamonds) and 20+1 (green circles) systems are in good agreement with the variational one particle-hole approximation of Ref.~\cite{Parish2021} (dashed blue line) and the the $T=0$ functional renormalization group results of Ref.~\cite{Punk2009} (dash-dotted orange line). Above $1/(k_F a) \sim 0.7$ the Monte Carlo sign problem becomes more severe and leads to large statistical errors in the AFMC calculations. Thus we cannot probe directly the polaron-molecule transition. The experimental results of Ref.~\cite{Ness2020}  are shown by the purple x's with error bars.

In the BEC regime of $(k_F a)^{-1} > 0$, there is a two-particle bound state whose binding energy  provides the leading contribution of $8\pi/ k_Fa$ to the contact (using Eq.~(\ref{eq:contactderivatives})). In Fig.~\ref{fig:Cvsa}(b) we show the contact with this dominant contribution removed as a function of $(k_F a)^{-1}$, and again compare with the variational one particle-hole approximation~\cite{Parish2021} and experiment~\cite{Ness2020}. We find that our results are once more consistent overall with the variational approximation. However, there are significant discrepancies between our AFMC results and the experimental results of Ref.~\cite{Ness2020}.  We also notice that around  $(k_F a)^{-1} \sim 0.5$ the dominant contribution from the two-particle binding energy overestimates the contact.\\

\subsection{Thermal energy gap}

\begin{figure}[b]
    \includegraphics[width=\columnwidth]{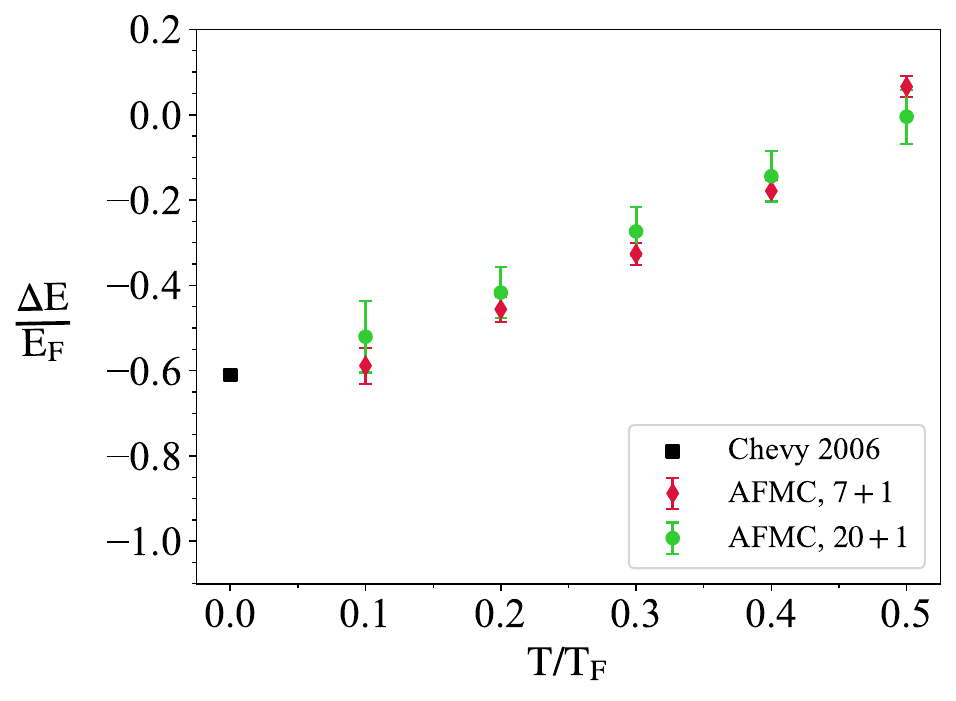}
    \caption{The thermal energy gap $\Delta E$ (in units of the Fermi energy $E_F$) as a function of temperature at unitarity. Our low-temperature results are consistent with the $T=0$ Chevy's ansatz of $\Delta E =-0.61 T_F$~\cite{Chevy2006}.}
    \label{fig:EGapvsT}
\end{figure}

We define the thermal energy gap to be $\Delta E = E(T) -E_{\rm NI}$, where $E=\langle \hat H\rangle$ is the thermal energy at temperature $T$ and $E_{\rm NI} $ is the non-interacting energy (i.e., the total energy when the contact interaction between the impurity and the medium is turned off).

In Fig.~\ref{fig:EGapvsT}, we show the energy gap (in units of the Fermi energy $E_F$) as a function of temperature for the 7+1 (red diamonds) and 20+1 (green circles) systems. We find that $\Delta E$ is monotonically increasing function of temperature in the regime shown in the figure.  Our results at low temperatures are consistent with the $T=0$ Chevy's ansatz of $\Delta E = -0.61\, E_F$~\cite{Chevy2006}.

\section{Conclusion and Outlook} 

We carried out the first controlled thermodynamic calculations for the Fermi polaron problem at strong coupling using canonical-ensemble AFMC methods on discrete lattices.  We eliminated systematic errors by extrapolating to continuous time and to the continuum limit.  The canonical-ensemble formulation, in which we project on fixed numbers of spin-up and spin-down particles, is particularly suitable for the Fermi polaron problem in that we project on $N_\downarrow=1$ (i,e, a single impurity) and $N_\uparrow$ particles in the Fermi sea. 

As a spin-imbalanced system, the Fermi polaron problem in general has a Monte Carlo sign problem, which usually represents a major obstacle in quantum Monte Carlo calculations. However, we found that there is a significant regime in parameter space (i.e., temperature and coupling strength $(k_F a)^{-1}$) in which the sign problem is moderate, enabling us to carry out precision Monte Carlo calculations. 

We studied the contact of the Fermi polaron system as a function of temperature at unitarity and as a function of the coupling strength at fixed temperature. We found good agreement with a variational method~\cite{Liu2020A,Parish2021} that employs a one particle-hole excitation in the Fermi sea. Our results for the contact vs.~temperature overall agree with the experiment of Ref.~\cite{Yan2019} considering the large experimental error bars, but for the dependence of the contact on the coupling strength we find significant deviations from the experiment of Ref.~\cite{Ness2020}.  

The dressed Fermi polaron is predicted to undergo a phase transition to a molecular dimer as a function of $(k_F a)^{-1}$. Currently we cannot probe directly the transition region because of the sign problem. However, the increase of the contact with temperature indicates the increased population of the excited molecular states~\cite{Liu2020A,Parish2021,Schmidt2011}.

We also calculated the thermal energy gap at unitarity as a function of temperature and found it to be consistent at low temperatures with Chevy's ansatz~\cite{Chevy2006} for the zero-temperature energy gap. 

In future work it would be interesting to calculate the polaron's spectral function using canonical-ensemble AFMC methods to gain insight into the quasiparticle  excitations of the polaron. 

\section{Acknowledgements}
We thank F. Chevy, J. Levinsen, N. Navon and M.M. Parish for helpful discussions. This work was supported in part by the U.S. DOE grants No.~DE-SC0019521 and No.~DE-SC0020177. The calculations used resources of the National Energy Research Scientific Computing Center (NERSC), a U.S. Department of Energy Office of Science User Facility operated under Contract No.~DE-AC02-05CH11231.  We thank the Yale Center for Research Computing for guidance and use of the research computing infrastructure.

\appendix 

\section{Continuous time and continuum extrapolations} 

\noindent \ssec{Extrapolations to continuous time} 
 The symmetric Trotter-Suzuki decomposition and the three-point Gaussian quadrature introduce a discretization error of order $(\Delta\beta)^2$. To obtain the  continuous time limit $\Delta\beta\rightarrow 0$, we calculate the relevant observables at different values of $\Delta\beta$ and carry out  a linear extrapolation in the dimensionless parameter $(\Delta\beta E_F)^2$ where $E_F$ is the Fermi energy.
  
\begin{figure}[tbh]
     \includegraphics[width=\columnwidth]{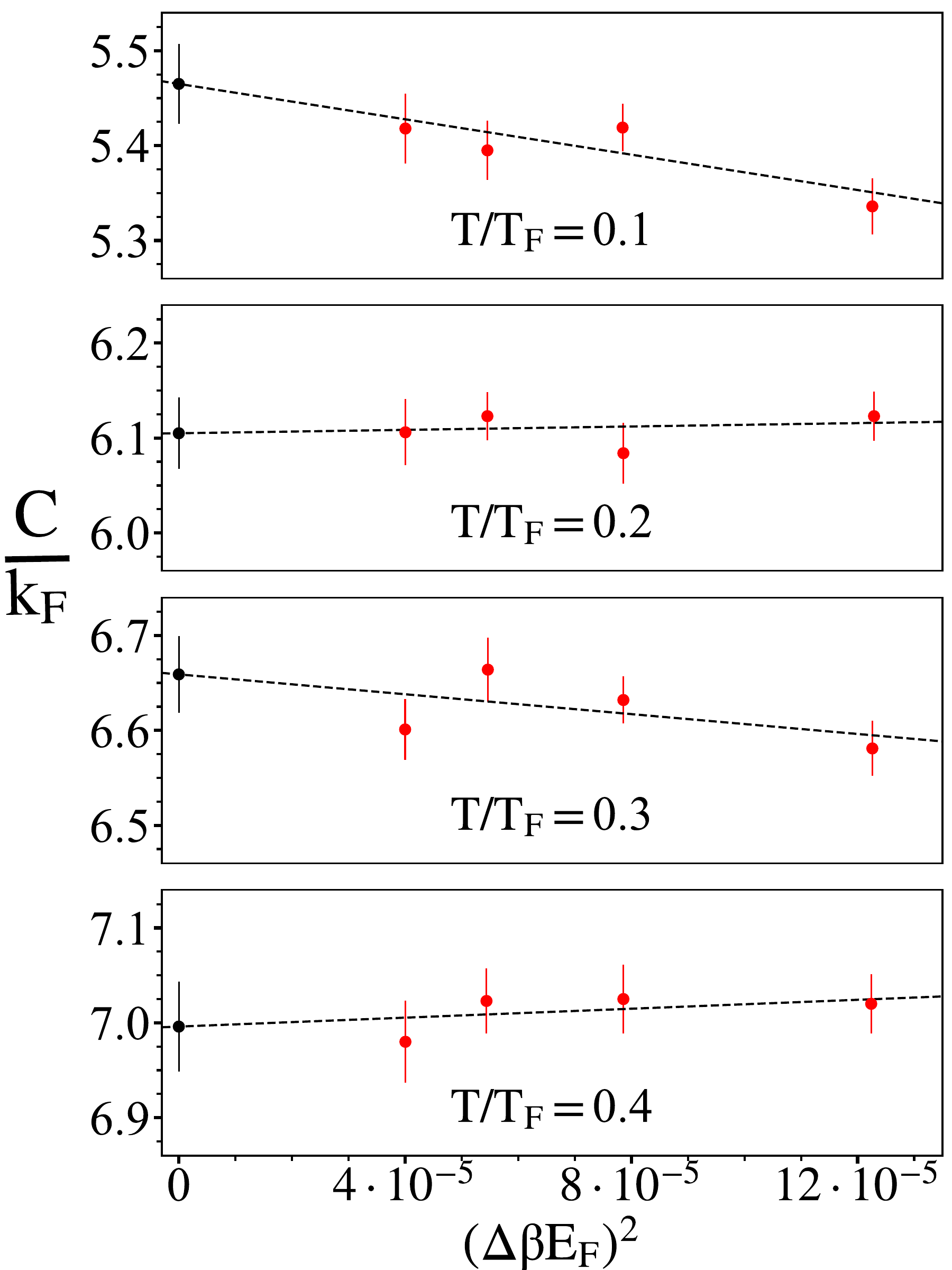}
    \caption{Examples of extrapolations to the continuous time  limit for $N=20+1$ particles at unitarity on a $9^3$ lattice. For sufficiently small time slices, the dependence on $\Delta\beta$ is less significant than the dependance on filling factor.}
    \label{fig:Contactdb}
\end{figure}

 We demonstrate the extrapolation to continuous time in Fig.~\ref{fig:Contactdb}, where we show the contact $C$ (in units of $k_F$) as a function of $(\Delta\beta E_F)^2$ for the $20+1$ system at unitarity on a $9^3$ lattice at several temperatures. The dashed lines are linear fits to the data (red circles) and the black circles are the extrapolated values of the contact at $\Delta\beta=0$.
 
For the thermal energy gap, the dependence on $\Delta\beta$ is smaller than the statistical fluctuations and we take an average (i.e., a flat line extrapolation).  \\

\noindent \ssec{Continuum extrapolations}
Following the $\Delta\beta \rightarrow 0$ extrapolation, we carry out the continuum extrapolation by calculating the observables on large lattices with filling factor $\nu=N_\uparrow/N_L^3 \rightarrow 0$ for fixed particle number $N_\uparrow$ in the Fermi sea and fixed temperature.  

\begin{figure}[htp]
    \includegraphics[width=\columnwidth]{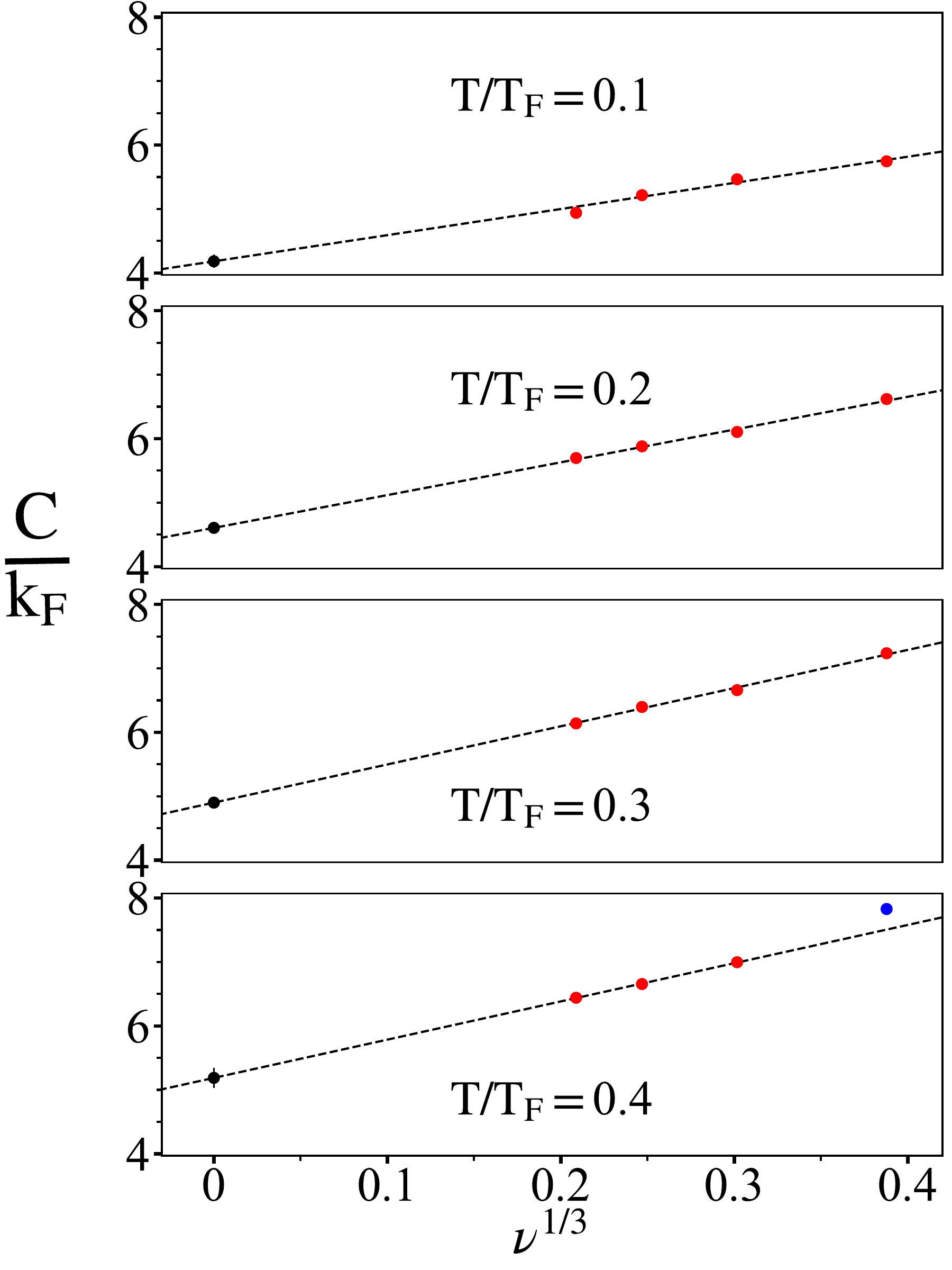}
    \caption{Examples of filling extrapolations for the contact for $N=20+1$ particles at unitarity. Linear fits are performed in $\nu^{1/3}$. We see a strong dependence of the contact on filling factor, as the contact decreases with decreasing filling factor.}
    \label{fig:ContactFilling}
\end{figure}

 In Fig.~\ref{fig:ContactFilling} we show the $\Delta\beta=0$ contact at unitarity for the $20+1$ system as a function of $\nu^{1/3}$ for several temperatures. The dashed lines are linear fits to the contact data (red circles) and the black circles are the extrapolated values of the contact at $\nu=0$.

\begin{figure}[htp]
    \centering
    \hspace*{-0.3cm}
    \includegraphics[scale=0.57]{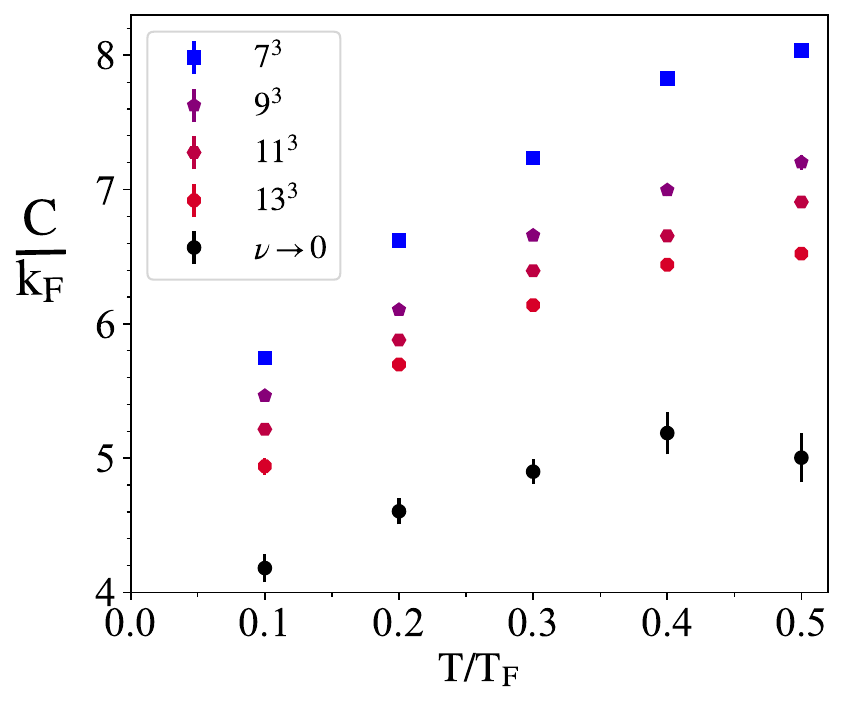}
    \caption{Contact as a function of temperature for $N=20+1$ particles at unitarity for different values of the filling factor $\nu$. The continuum limit results are shown in black; we find that the continuum limit results defer significantly from the results obtained at finite filling factor.}
    \label{fig:continuum}
\end{figure}

The continuum extrapolations are important~\cite{Jensen2020} in that they can modify the behavior of observables even on a qualitative level. An example is shown in Fig.~\ref{fig:continuum} for the contact at unitarity for the $20+1$ system as a function of temperature.  The contact is shown for different values of the filling factor $\nu$ (i.e., different lattice sizes) and the extrapolated contact in the continuum limit $\nu=0$ is shown by the black circles. We see the extrapolated contact increases more slowly with temperature when compared to the contact at a given lattice size.  Our calculations for large lattice sizes were made possible  by the controlled model space truncation algorithm introduced in Ref.~\cite{Gilbreth2021}. 

\end{document}